\documentclass[aps,prd,twocolumn,showpacs,floats,floatfix,letterpaper,nofootinbib,superscriptaddress,]{revtex4}

\usepackage{amssymb,amsmath,latexsym,mathrsfs}
\usepackage{graphicx}
\usepackage{epsfig}
\usepackage{multirow}
\usepackage{array}
\usepackage{xspace}
\usepackage{color}

\newcommand{\bea}{\begin{eqnarray}}
\newcommand{\eea}{\end{eqnarray}}

\newcommand{\neff}{N_{\textrm{eff}}}

\newcommand{\e}[1]{\times 10^{#1}}
\newcommand{\mpcinv}{\, \text{Mpc}^{-1}}

\newcommand{\pchip}{\texttt{PCHIP}\xspace}

\begin{document}


\title{Robustness of cosmological axion mass limits}

\author{Eleonora Di Valentino}
\affiliation{CNRS, UMR 7095, Institut d'Astrophysique de Paris, F-75014, Paris, France}
\affiliation{Sorbonne Universit\'es, Institut Lagrange de Paris (ILP), F-75014, Paris, France}
\author{Stefano Gariazzo}
\affiliation{Department of Physics, University of Torino, Via P. Giuria 1, I--10125 Torino, Italy}
\affiliation{INFN, Sezione di Torino, Via P. Giuria 1, I--10125 Torino, Italy}
\author{Elena Giusarma}
\affiliation{Physics Department and INFN, Universit\`a di Roma ``La Sapienza'', Ple Aldo Moro 2, 00185, Rome, Italy}
\author{Olga Mena}
\affiliation{IFIC, Universidad de Valencia-CSIC, 46071, Valencia, Spain}
\begin{abstract}

We present cosmological bounds on the thermal axion mass in an extended cosmological scenario in which the primordial power
spectrum of scalar perturbations differs from the usual power-law shape predicted by the simplest inflationary models. The power
spectrum is instead modeled by means of a "piecewise cubic Hermite interpolating polynomial'' (\pchip). 
When using Cosmic Microwave Background measurements combined with other cosmological data sets, the thermal axion mass constraints are degraded only slightly. The addition of the measurements of $\sigma_8$ and $\Omega_m$ from the 2013 Planck cluster catalogue on galaxy number counts relaxes the bounds on the thermal axion mass, mildly favouring a $\sim 1$~eV axion mass, regardless of the model adopted for the primordial power spectrum. However, in general, such a preference disappears if the sum of the three active neutrino masses is also considered as a free parameter in our numerical analyses, due to the strong correlation between the masses of these two hot thermal relics. 
\end{abstract}


\maketitle

\section{Introduction}
\label{sec:intro}
A possible candidate for an extra hot thermal relic component is the
axion particle produced thermally in the early universe. Axions
therefore can contribute to the hot dark matter component together
with the standard relic neutrino background. 
Axions may be produced in the early universe via thermal or non
thermal processes, and arise as the solution to solve the strong CP
problem~\cite{PecceiQuinn,Weinberg:1977ma,Wilczek:1977pj}. Axions are the Pseudo-Nambu-Goldstone
bosons of a new global $U(1)_{PQ}$ (Peccei-Quinn) symmetry that is 
spontaneously broken at an energy scale $f_a$.
 The axion mass is given by
\begin{equation}
m_a = \frac{f_\pi m_\pi}{  f_a  } \frac{\sqrt{R}}{1 + R}=
0.6\ {\rm eV}\ \frac{10^7\, {\rm GeV}}{f_a}~,
\end{equation}
where $f_a$ is the axion coupling constant, $R=0.553 \pm 0.043 $ is the up-to-down quark masses
ratio and $f_\pi = 93$ MeV is the pion decay constant. Non-thermal axions, as those produced by the misalignment mechanism, 
while being a negligible hot dark matter candidate, may constitute a fraction or the total cold dark matter component of the universe.
We do not explore such a possibility here. 
Thermal axions will affect the cosmological observables in a very
similar way to that induced by the presence of neutrino masses and/or
extra sterile neutrino species. 
Massive thermal axions as hot relics affect large scale structure,
since they will only cluster at scales larger than
their free-streaming scale when they become non-relativistic,
suppressing therefore structure formation at small scales (large
wavenumbers $k$). 
Concerning Cosmic Microwave Background (CMB) physics, 
an axion mass will also lead to a signature in the CMB photon
temperature anisotropies via the early integrated Sachs-Wolfe effect.
In addition, extra light species as thermal axions
will contribute to the dark radiation content of the universe, or, in
other words, will lead to an increase of the effective number of relativistic degrees of freedom $\neff$, defined via
\begin{equation}
 \rho_{rad} = \left[1 + \frac{7}{8} \left(\frac{4}{11}\right)^{4/3}\neff\right]\rho_{\gamma} \, ,
\end{equation}
where $\rho_{\gamma}$ refers to the present photon energy density. 
In the standard cosmological model in which a thermal axion content is
absent, the three active neutrino contribution leads to the canonical
value of $\neff=3.046$ \cite{Mangano:2005cc}. The extra contribution
to $\neff$ arising from thermal axions can modify both the
CMB anisotropies (via Silk damping) and
the light element primordial abundances predicted by Big Bang
Nucleosynthesis. 
The former cosmological signatures of thermal axions have been extensively
exploited in the literature to derive bounds on the thermal axion
mass, see Refs.~\cite{Melchiorri:2007cd,Hannestad:2007dd,Hannestad:2008js,Hannestad:2010yi,Archidiacono:2013cha,Giusarma:2014zza}.

However, all the cosmological axion mass limits to date have assumed the usual simple power-law
description for the primordial perturbations.  The aim of this paper
is to constrain the mass of the thermal axion using  a non-parametric
description of the  Primordial Power Spectrum (PPS hereinafter) of the
scalar perturbations, as introduced in Ref.~\cite{Gariazzo:2014dla}.
While in the simplest models of
inflation~\cite{
Guth:1980zm,Linde:1981mu,Starobinsky:1982ee,Hawking:1982cz,Albrecht:1982wi,Mukhanov:1990me,Mukhanov:1981xt,Lucchin:1984yf, Lyth:1998xn,Bassett:2005xm,Baumann:2008bn}
the PPS has a scale-free power-law form, the PPS could be more complicated, presenting various features or a scale dependence.
Several methods have been proposed in the literature to reconstruct
the shape of the PPS (see the recent  work of Ref.~\cite{Ade:2015lrj}). It has been
shown~\cite{Hunt:2013bha,Hazra:2014jwa} that there are small hints for deviations from the power-law form,
even when using different methods and different data sets.

The energy scales at which the PPS was produced during inflation
can not be directly tested. We can only infer the PPS by measuring the
current matter power spectrum in the galaxy distribution and the power
spectrum of the CMB fluctuations. The latter one, measured with exquisite 
precision by the Planck experiment \cite{Planck:2015xua,Ade:2013zuv,Ade:2013ktc}, 
is the convolution of the PPS with the transfer function. 
Therefore, in order to reconstruct the PPS, the assumption of an underlying cosmological
model is a mandatory first step in order to compute the transfer function.

Here we rather exploit a non-standard PPS approach, which can allow for a good fit to
experimental data even in models that deviates from the standard
cosmological picture. In particular, we consider a thermal axion
scenario, allowing the PPS to assume a more general shape than the
usual power-law description. This will allow us to test the robustness of the cosmological thermal axion mass bounds (see Ref.~\cite{Giusarma:2014zza} for a recent standard thermal axion analysis), as first performed in Ref.~\cite{dePutter:2014hza} for the neutrino mass case.  

The structure of the paper is as follows. Section \ref{sec:method}
describes the PPS modeling used in this study, as well as the
description of the thermal axion model explored here and the
cosmological data sets exploited to constrain such a model. In
Sec. \ref{sec:results} we present and discuss the results arising from
our bayesian analysis, performed through the Monte Carlo Markov Chains (MCMC) package \texttt{CosmoMC} \cite{Lewis:2002ah}, 
while the calculation of the theoretical observables is done through the Boltzman equations solver 
\texttt{CAMB} (Code for Anisotropies in the Microwave Background) \cite{Lewis:1999bs}.
We draw our conclusions in Sec.~\ref{sec:concl}.

\section{Method}
\label{sec:method}
In this section we focus on the tools used in the numerical analyses performed here.
Subsection~\ref{sub:pps} describes the alternative model for the PPS
of scalar perturbations used for the analyses here (see also Ref.~\cite{Gariazzo:2014dla}),
while in Subsection~\ref{sub:model} we introduce the cosmological
model and the thermal axion treatment followed in this study. Finally,
we shall present in Subsection~\ref{sub:data}
 the cosmological data sets used in the MCMC analyses.

\subsection{Primordial Power Spectrum Model}
\label{sub:pps}
The primordial fluctuations in scalar and tensor modes are generated during the inflationary phase in the early universe.
The simplest models of inflation predict a power-law form for the PPS of scalar and tensor perturbations 
(see e.g. ~\cite{Guth:1980zm,Linde:1981mu,Starobinsky:1982ee,Hawking:1982cz,Albrecht:1982wi,Mukhanov:1990me,Mukhanov:1981xt,Lucchin:1984yf,Lyth:1998xn,Bassett:2005xm,Baumann:2008bn}
and references therein),
but in principle inflation can be generated by more complicated mechanisms, thus giving a different shape for the PPS 
(see Refs.~\cite{Martin:2014vha,Kitazawa:2014dya} and references therein). In order to study how the cosmological constraints on the parameters change in more general inflationary scenarios, we assume a non-parametric form for the PPS.

Among the large number of possibilities, we decided to describe the
PPS of scalar perturbations using a function to interpolate the PPS
values in a series of nodes at fixed position. The interpolating
function we used is named ``piecewise cubic Hermite interpolating
polynomial'' (\pchip) \cite{Fritsch:1980} and it is a modified spline
function, defined to preserve the original monotonicity of the point series that is interpolated.
We use a modified version of the original \pchip algorithm \cite{Fritsch:1984},  detailed in Appendix~A of Ref.~\cite{Gariazzo:2014dla}.

To describe the scalar PPS with the \pchip model, we only need to give the values of the PPS in a discrete number of nodes and to interpolate among them.
We use 12 nodes
which span a wide range of $k$-values:
\begin{align}
k_1     &= 5\e{-6} \mpcinv , \nonumber\\
k_2     &= 10^{-3} \mpcinv , \nonumber\\
k_j     &= k_2 (k_{11}/k_2)^{(j-2)/9} \quad \text{for} \quad j\in[3,10] , \nonumber\\
k_{11}  &= 0.35 \mpcinv , \nonumber\\
k_{12}  &= 10\mpcinv .
\label{eq:nodesspacing}
\end{align}
We choose equally spaced nodes in the logarithmic scale in the range $(k_2, k_{11})$, that is well constrained from the data \cite{dePutter:2014hza},
while the first and the last nodes are useful to allow for a non-constant behaviour of the PPS outside the well-constrained range.

The \pchip PPS is described by
\begin{equation}
P_{s}(k)=P_0 \times \pchip(k; P_{s,1}, \ldots, P_{s,12})
,
\label{eq:pchip}
\end{equation}
where $P_{s,j}$ is the value of the PPS at the node $k_j$ divided by $P_0=2.36\e{-9}$ \cite{Larson:2010gs}.
\subsection{Cosmological and Axion Model}
\label{sub:model}

The baseline scenario we consider here is the $\Lambda$CDM model, extended with hot thermal relics (the axions), together
with the PPS approach outlined in the previous section. For the numerical analyses 
we use the following set of parameters, for which we assume flat priors in the intervals listed in Tab.~\ref{tab:priors}:
\begin{equation}\label{parameterPPS}
\{\omega_b,\omega_c, \Theta_s, \tau, m_a, \sum m_\nu, P_{s,1}, \ldots, P_{s,12}\}~,
\end{equation}
where $\omega_b\equiv\Omega_bh^{2}$ and $\omega_c\equiv\Omega_ch^{2}$  
are, respectively, the physical baryon and cold dark matter energy densities,
$\Theta_{s}$ is the ratio between the sound horizon and the angular
diameter distance at decoupling, $\tau$ is the reionization optical depth, $m_a$ and $\sum m_\nu$ are the axion and the sum of three active neutrino masses (both in eV) and $P_{s,1}, \ldots,
P_{s,12}$ are the parameters of the \pchip PPS. We shall also consider a scenario in which massive neutrinos are also present, to explore the expected degeneracy between the sum of the neutrino masses and the thermal axion mass~\cite{Giusarma:2014zza}. 

In order to compare the results obtained with the \pchip PPS to the results obtained with the usual power-law PPS model, we describe the latter case with the following set of parameters:
\begin{equation}\label{parameterPL}
\{\omega_b,\omega_c, \Theta_s, \tau, m_a, n_s, \log[10^{10}A_{s}]\}~,
\end{equation}
where $n_s$ is the scalar spectral index, $A_{s}$ the amplitude of the primordial spectrum
and the other parameters are the same ones described above. The case of several hot thermal relics for the standard scenario will not be carried out here, as it has been done in the past by several authors (see e.g.~\cite{Giusarma:2014zza}).
The flat priors we use are listed in Tab.~\ref{tab:priors}.

\begin{table}
\begin{center}
\begin{tabular}{c|c}
Parameter                    & Prior\\
\hline
$\Omega_{\rm b} h^2$         & $[0.005,0.1]$\\
$\Omega_{\rm cdm} h^2$       & $[0.001,0.99]$\\
$\Theta_{\rm s}$             & $[0.5,10]$\\
$\tau$                       & $[0.01,0.8]$\\
$m_a$ (eV)                        & $[0.1,3]$\\
$\sum m_\nu$ (eV)               & $[0.06,3]$\\
$P_{s,1}, \ldots, P_{s,12}$  & $[0.01,10]$\\
$n_s$                        & $[0.9, 1.1]$\\
$\log[10^{10}A_{s}]$         & $[2.7,4]$\\
\end{tabular}
\end{center}
\caption{
Priors for the parameters used in the MCMC analyses.
}
\label{tab:priors}
\end{table}

\begin{figure*}[!t]
\includegraphics[width=15cm]{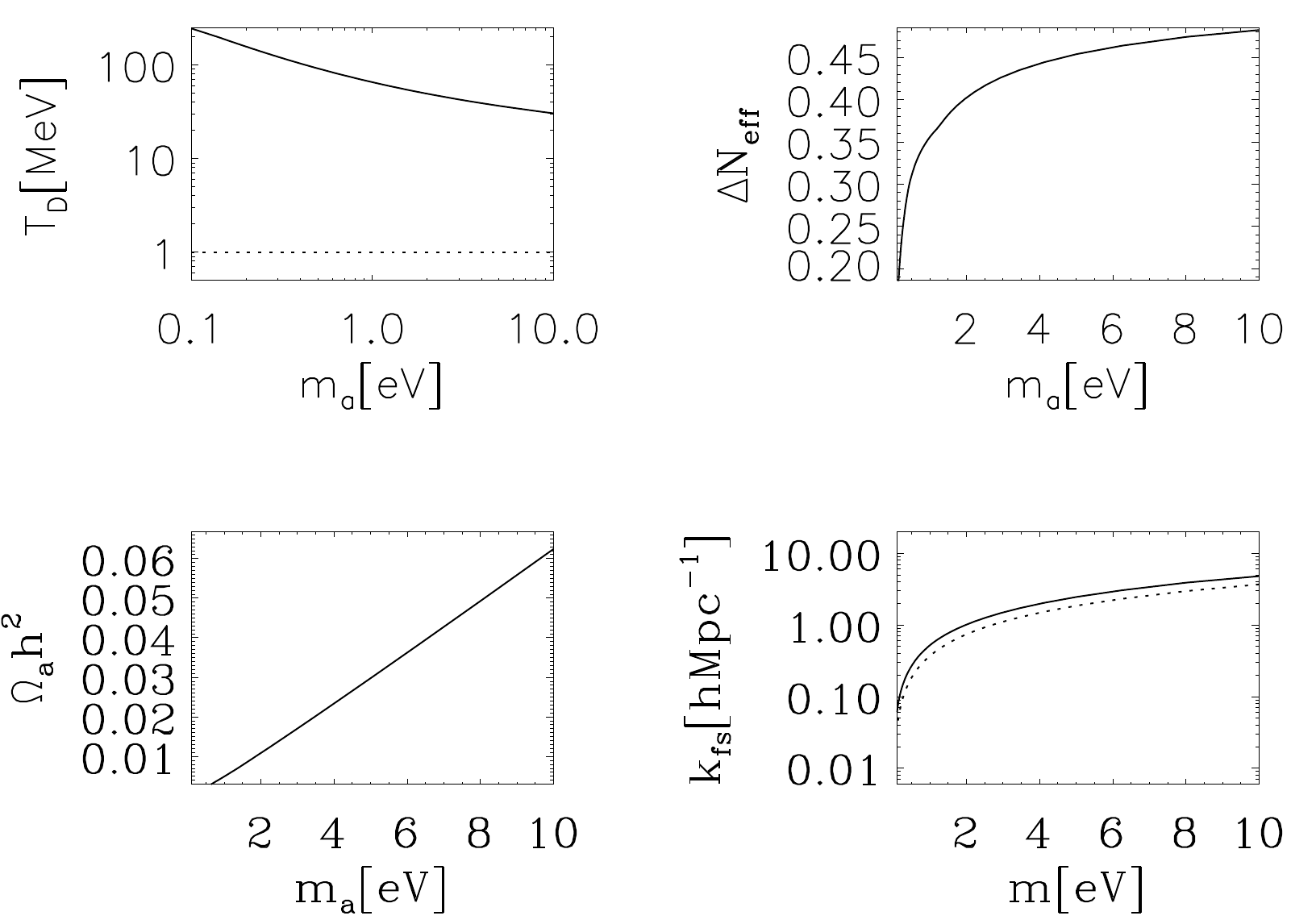}
\caption{The left upper panel shows the temperature of decoupling as a function of the axion mass (solid curve), as well as the Big Bang Nucleosynthesis temperature, $T_{\textrm{BBN}}\simeq 1$~MeV (dashed curve). The right upper panel shows the axion contribution to the extra dark radiation content of the universe, while the bottom right plot depicts the free-streaming scale of an axion (solid curve) or a neutrino (dashed curve) versus the axion/neutrino mass, in eV. The left bottom panel shows the current axion mass-energy density as a function of the axion mass.}
\label{fig:maref}
\end{figure*}

Concerning the contribution of the axion mass-energy density to
the universe's expansion rate, we briefly summarize our treatment in the following.
Axions decoupled in the early universe at a temperature $T_D$ given by
the  usual freeze out condition for a thermal relic:
\bea
\Gamma (T_D) = H (T_D)~,
\label{eq:decouplinga}
\eea 
where the thermally averaged interaction rate $\Gamma$ refers to the $\pi + \pi \rightarrow \pi
+a$ process:
\bea
\Gamma = \frac{3}{1024\pi^5}\frac{1}{f_a^2f_{\pi}^2}C_{a\pi}^2 I_a~,
\eea
with $C_{a\pi} = \frac{1-R}{3(1+R)}$ representing the axion-pion coupling constant  and the integral $I_a$ reads as follows
\bea
I_a &=&n_a^{-1}T^8\int dx_1dx_2\frac{x_1^2x_2^2}{y_1y_2}
f(y_1)f(y_2) \nonumber \\
&\times&\int^{1}_{-1}
d\omega\frac{(s-m_{\pi}^2)^3(5s-2m_{\pi}^2)}{s^2T^4}~,
\eea
in which $n_a=(\zeta_{3}/\pi^2) T^3$ refers to the number density for axions in thermal equilibrium. The function $f(y)=1/(e^y-1)$ is the pion thermal distribution and there are three different kinematical variables ($x_i=|\vec{p}_i|/T$,  $y_i=E_i/T$ ($i=1,2$) and $s=2(m_{\pi}^2+T^2(y_1y_2-x_1x_2\omega))$). 
The freeze out equation above, Eq.~(\ref{eq:decouplinga}), can be numerically solved, obtaining the axion decoupling temperature
$T_D$ as a function of the axion mass $m_a$. Figure~\ref{fig:maref} shows, in the left upper panel, the axion decoupling temperature as a function of the axion mass, in eV units. Notice that, the higher the axion mass, the lower  the temperature of decoupling is. From the axion decoupling temperature it is possible to infer the present axion number density,
related to the current photon density $n_\gamma$ by 
\bea
n_a=\frac{g_{\star S}(T_0)}{g_{\star S}(T_D)} \times \frac{n_\gamma}{2}~, 
\label{eq:numberdens}
\eea  
where $g_{\star S}$ represents the number of \emph{entropic} degrees of
freedom, with $g_{\star S}(T_0) = 3.91$. The contribution of the relic axion to the total mass-energy density of the universe will be given by the product of the axion mass times the axion number density. The quantity $\Omega_a h^2$ at the present epoch is depicted in the bottom left panel of Fig.~\ref{fig:maref}. Notice that, currently, a $1$~eV axion will give rise to $\Omega_a h^2\simeq 0.005$, while a neutrino of the same mass will contribute to the total mass-energy density of the universe with $\Omega_\nu h^2\simeq 0.01$. Notice however that  $\Omega_a h^2$ represents the contribution from relic, thermal axion states. Non-thermal processes, as the misalignment production, could also produce a non-thermal axion population which we do not consider here, see the work  of \cite{DiValentino:2014zna} for the most recent cosmological constraints on such scenario. As previously stated, the
presence of a thermal axion will also imply an extra radiation component at the BBN period:
\begin{equation}
\Delta \neff =\frac{ 4}{7}\left(\frac{3}{2}\frac{n_a}{n_\nu}\right)^{4/3}~,
\end{equation}
where $n_a$ is given by Eq.~(\ref{eq:numberdens}) and $n_\nu$ refers
to the present neutrino plus antineutrino number density per flavour. The top right panel of Fig.~\ref{fig:maref} shows the axion contribution to the radiation component of the universe as a function of the axion mass. Notice that the extra dark radiation arising from a $1$~eV axion is still compatible (at $95\%$~CL) with the most recent measurements of $\neff$ from the Planck mission~\cite{Planck:2015xua}. The last crucial cosmological axion quantity is the axion free streaming scale, i.e. the wavenumber $k_{\rm {fs}}$ below which axion density perturbations will contribute to clustering once the axion is a non-relativistic particle. This scale is illustrated in Fig.~\ref{fig:maref}, in the bottom right panel, together with that corresponding to a neutrino of the same mass. Notice that they cover the same scales for our choice of priors for $m_a$ and $\sum m_\nu$ and therefore one can expect a large correlation between these two quantities in measurements of galaxy clustering. We will explore this degeneracy in the following sections. We summarize the axion parameters in Tab.~\ref{tab:axionparams}, where we specify the values of the decoupling temperature, $\Delta \neff$, $\Omega_a h^2$ and $k_{\rm {fs}}$ for the range of axion masses considered here, $(0.1, 3)$~eV.

\begin{table}
\begin{center}
\begin{tabular}{|c|c|c|}
\hline
Axion parameter                    &  & \\
\hline\hline
$m_a$ (eV)                        & $0.1$ &$3$ \\
$T_D$  (MeV)       & $245.6$ &$43.2$ \\
$\Omega_a h^2$       & $0.0003$ &$0.016$ \\
$\Delta \neff$            & $0.18$ &$0.43$ \\          
$k_{\rm {fs}}$ ($h$/Mpc) &$0.06$ &$1.46$\\
\hline
\end{tabular}

\end{center}
\caption{Values for the axion parameters, $T_D$, $\Delta \neff$, $\Omega_a h^2$ and $k_{\rm {fs}}$ for the lower and upper prior choice of $m_a$ explored here. }
\label{tab:axionparams}
\end{table}

\subsection{Cosmological measurements}
\label{sub:data}

Our baseline data set consists of CMB measurements. These include the
temperature data from the Planck satellite, see
Refs.~\cite{Ade:2013ktc,Planck:2013kta}, together with the
WMAP 9-year polarization measurements, following
\cite{Bennett:2012fp}. We also consider high multipole data from the
South Pole Telescope (SPT) \cite{Reichardt:2011yv} as well as from the Atacama Cosmology Telescope (ACT) \cite{Das:2013zf}. 
The combination of all the above CMB data  is referred to as the \emph{CMB} data set.

\begin{figure*}[!t]
\begin{center}
\includegraphics[width=12.cm]{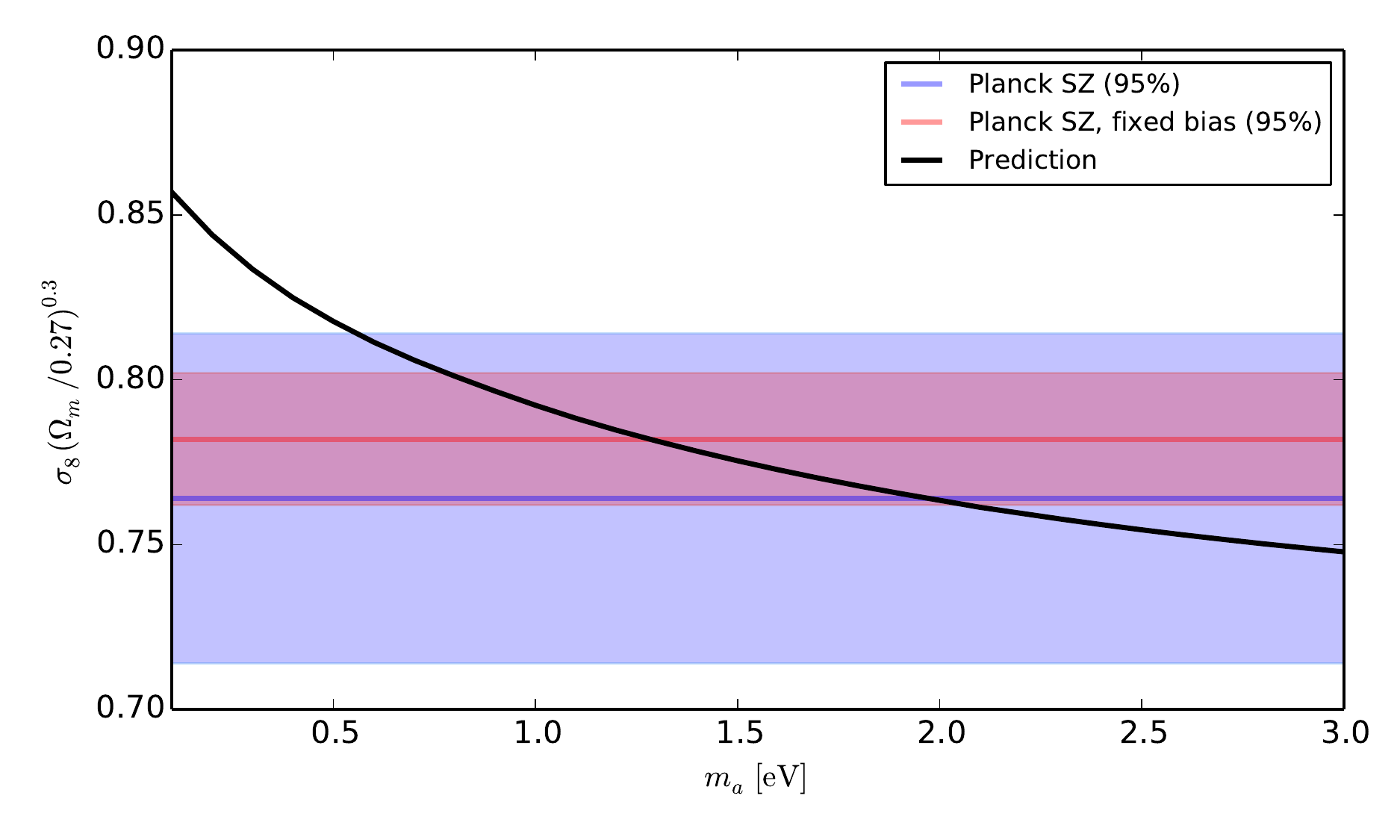}
\end{center}
 \caption{Cluster normalisation condition, $\sigma_8
(\Omega_m/0.27)^{0.3}$, as a function of the thermal axion mass. We also show the current PSZ measurements with their associated $95\%$~CL uncertainties.}
\label{fig:sigma8}
\end{figure*}

Galaxy clusters represent an independent tool to probe the cosmological
parameters.  Cluster surveys usually report their measurements by means
 of the so-called cluster normalization condition, $\sigma_8
 \Omega^\gamma_m$, where $\gamma \sim 
 0.4$~\cite{Allen:2011zs,Weinberg:2012es,Rozo:2013hha}.
We shall use here the cluster normalization condition as measured by
the Planck Sunyaev-Zeldovich (PSZ) 2013 catalogue~\cite{Ade:2013lmv},
referring to it as the \emph{PSZ} data set. The PSZ measurements of
the cluster mass function provide the constraint $\sigma_8
(\Omega_m/0.27)^{0.3}=0.764\pm 0.025$. 
As there exists a strong degeneracy between the value of the
$\sigma_8$ parameter and the cluster mass bias, it is possible to fix the value of the
bias parameter accordingly to the results arising from numerical
simulations. In this last case, the error on the cluster
normalization condition from the PSZ catalogue is considerably reduced:
$\sigma_8 (\Omega_m/0.27)^{0.3}=0.78\pm 0.01$. In our analyses, we shall consider the two PSZ measurements of the cluster normalization condition, to illustrate the impact of the cluster mass bias in the thermal axion mass bounds, as recently explored in Ref.~\cite{Ade:2015fva} for the neutrino mass case. Figure \ref{fig:sigma8} illustrates the prediction for the cluster normalisation condition, $\sigma_8
(\Omega_m/0.27)^{0.3}$, as a function of the thermal axion mass. We also show the current PSZ measurements with their associated $95\%$~CL uncertainties, including those in which the cluster mass bias parameter is fixed. Notice that the normalisation condition decreases as the axion mass increases, due to the decrease induced in the $\sigma_8$ parameter in the presence of axion masses: the larger the axion mass is, the larger the reduction in the matter power spectra will be. 

Tomographic weak lensing surveys are sensitive to the overall
amplitude of the matter power spectrum by measuring the correlations in the observed shape of distant 
galaxies induced by the intervening large scale structure.  
The matter power spectrum amplitude depends on both the $\sigma_8$
clustering parameter and the matter density $\Omega_m$. Consequently, tomographic lensing surveys, via
measurements of the galaxy power shear spectra, provide additional
and independent constraints in the ($\sigma_8$, $\Omega_m$) plane. 
The Canada-France-Hawaii Telescope Lensing Survey, CFHTLenS, with six tomographic redshift bins (from $z=0.28$ to
$z=1.12$), provides a constraint on the relationship between $\sigma_8$
and $\Omega_m$ of $\sigma_8 (\Omega_m/0.27)^{0.46}=0.774\pm
0.040$~\cite{Heymans:2013fya}. We shall refer to this data set as \emph{CFHT}.

We also address here the impact of a gaussian prior on the Hubble constant
$H_0=70.6\pm3.3$ km/s/Mpc from an independent reanalysis of Cepheid
data~\cite{Efstathiou:2013via}, referring to this prior as the 
\emph{HST} data set. 

We have also included measurements of the large scale structure of the
universe in their geometrical form, i.e., in the form of Baryon Acoustic
Oscillations (BAO). Although previous studies in the literature have shown that, 
for constraining hot thermal relics, the shape information contained in the galaxy power spectrum is more
powerful when dealing simultaneously with extra relativistic species
and hot thermal relic masses~\cite{Hamann:2010pw,Giusarma:2012ph}, we exploit here the BAO signature, as the
contribution from the thermal axions to the relativistic number of
species is not very large (see Tab.~\ref{tab:axionparams}), and current measurements from galaxy surveys are
mostly reported in the geometrical (BAO) form.

\begin{figure*}[!t]
\includegraphics[width=10.cm]{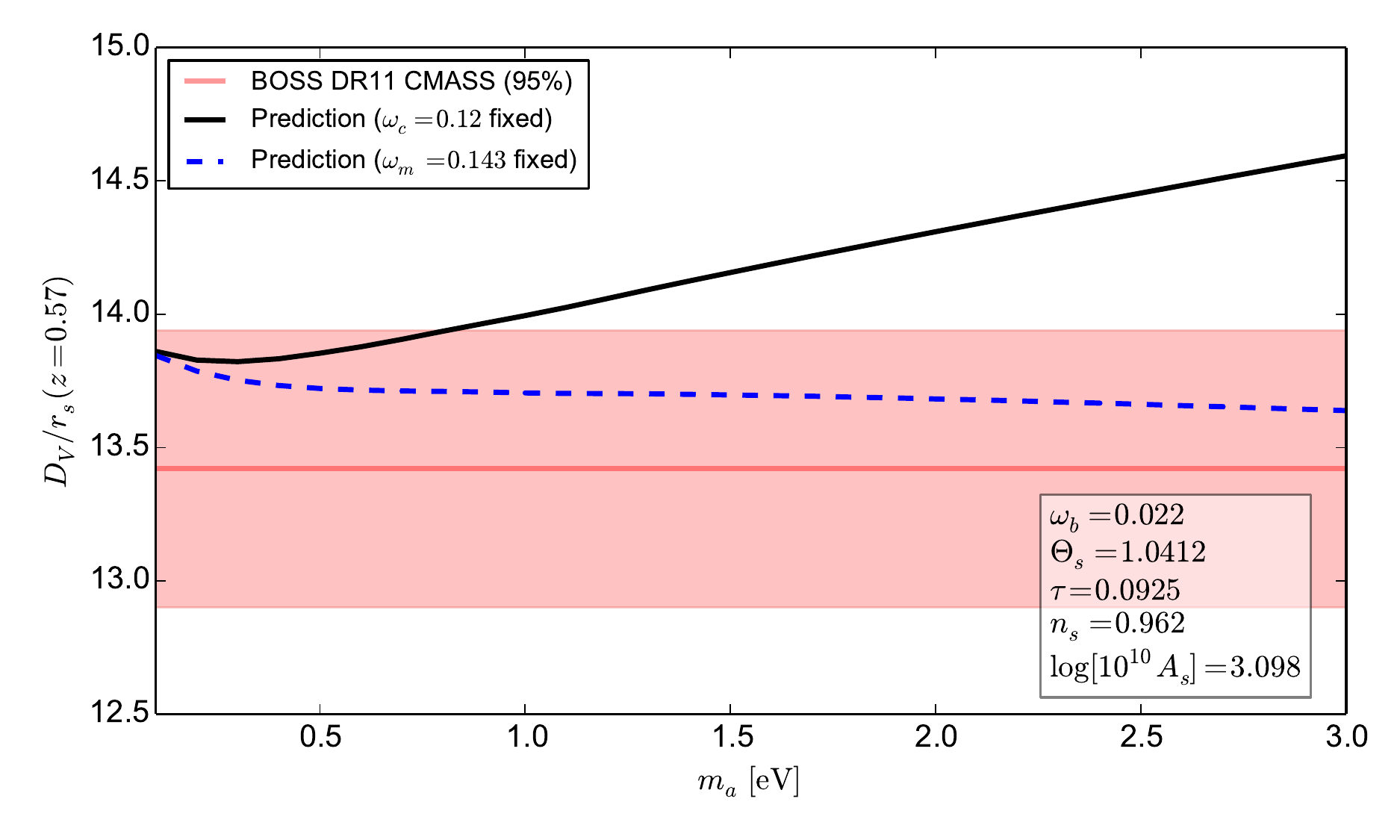}
 \caption{The solid black line depicts the spherically averaged BAO distance $D_V(z)$, as a function of the axion mass, at a redshift of $z=0.57$, after keeping fixed all the remaining cosmological parameters, the cold dark matter included. The dashed blue line depicts the equivalent but keeping fixed the total matter mass-energy density (and consequently changing the cold dark matter $\omega_c$). The bands show the measurement from the BOSS experiment (DR11) with its associated $95\%$~CL error.}
\label{fig:dvboss}
\end{figure*}

The BAO wiggles, imprinted in the power spectrum of the galaxy
distribution, result from the competition in the coupled photon-baryon
fluid between radiation pressure and gravity. The BAO measurements
that have been considered in our numerical analyses include the
results from the WiggleZ~\cite{Blake:2011en}, the
6dF~\cite{Beutler:2011hx} and the SDSS II surveys~\cite{Percival:2009xn,Padmanabhan:2012hf}, at redshifts of  $z=0.44,
0.6, 0.73$, $z=0.106$ and $z=0.35$, respectively. We also include in our analyses as well the Data Release 11
(DR11) BAO signal of the BOSS experiment~\cite{Dawson:2012va}, which provides the most precise distant
constraints~\cite{Anderson:2013zyy} measuring both the Hubble parameter and the angular diameter distance at an effective redshift of $0.57$. 
Figure \ref{fig:dvboss} illustrates the spherically averaged BAO distance, $D_V(z) \propto D^2_A(z)/H(z)$, as a function of the axion mass, at a redshift of $z=0.57$, as well as the measurement from the BOSS experiment with $95\%$~CL error bars~\cite{Anderson:2013zyy}. Notice that, from background measurements only, there exists a strong degeneracy between the cold dark matter mass-energy density and the axion one. The solid black line in Fig.~ \ref{fig:dvboss} shows the spherically averaged BAO distance if all the cosmological parameters are fixed, including $\omega_c$. The spherically averaged BAO distance deviates strongly from the $\Lambda$CDM prediction. However,  if $\omega_c$ is varied while $m_a$ is changed (in order to keep the total matter mass-energy density constant, see the dotted blue line in Fig.~\ref{fig:dvboss}), the  spherically averaged BAO distance approaches to its expected value in a $\Lambda$CDM cosmology.

\subsection{Compatibility of data}
It has been pointed out (see Sec. 5.5 of Ref.~\cite{Ade:2013zuv} and also Refs.~\cite{Giusarma:2014zza,Leistedt:2014sia}) that the value of $\sigma_8$ reported by cluster measurements and the value estimated from Planck CMB  measurements show a tension at the $\sim 2\sigma$ level. These discrepancies may arise due to the lack of a full understanding of the cluster mass calibrations. Although some studies in the literature, including the present one, show that in extended cosmological models with non-zero neutrino masses the discrepancies previously mentioned could be alleviated, the results from Ref.~\cite{Leistedt:2014sia} show, using also Bayesian evidence, that a canonical $\Lambda$CDM scenario with no massive neutrinos is preferred over its neutrino extensions by several combinations of cosmological datasets. Therefore, the results presented here and obtained when considering cluster data depend strongly on the reliability of low-redshift cluster data. If future data confirm current low-redshift cluster measurements, one could further test some of the possible beyond the $\Lambda$CDM models using particle physics experiments. For instance, the existence of a full thermal sterile neutrino  could be tested with neutrino oscillation experiments, and the active neutrino mass could also be tested by tritium experiments or, if the neutrino is a Majorana particle, by neutrinoless double beta decay searches.

\section{Results}
\label{sec:results}

\begin{table*}
\begin{center}\footnotesize
\begin{tabular}{lcccccccc}
\hline \hline
                         & CMB                         & CMB+HST                     & CMB+BAO                     & CMB+BAO                     & CMB+BAO  & CMB+BAO                                         & CMB+BAO\\
                         &                             &                             &                             &    +HST                     & HST+CFHT &  +HST+PSZ  (fixed bias)                                     &  +HST+PSZ\\  
\hline
\hspace{1mm}\\

$\Omega_{\textrm{c}}h^2$ & $0.127\,^{+0.007}_{-0.007}$ & $0.122\,^{+0.006}_{-0.006}$ & $0.122\,^{+0.003}_{-0.003}$ & $0.121\,^{+0.003}_{-0.003}$ & $0.120\,^{+0.003}_{-0.003}$ & $0.118\,^{+0.002}_{-0.002}$ & $0.119\,^{+0.003}_{-0.004}$ \\
\hspace{1mm}\\

$m_a$ [eV]               & {\rm {Unconstrained}}       & $<1.31$                     & $<0.89$                     & $<0.91$                     & $<1.29$                     & $1.00\,^{+0.50}_{-0.48}$    & $0.93\,^{+0.70}_{-0.71}$    \\
\hspace{1mm}\\

$\sigma_8$               & $0.788\,^{+0.079}_{-0.086}$ & $0.821\,^{+0.052}_{-0.074}$ & $0.827\,^{+0.044}_{-0.057}$ & $0.825\,^{+0.045}_{-0.059}$ & $0.793\,^{+0.049}_{-0.058}$ & $0.760\,^{+0.023}_{-0.022}$ & $0.767\,^{+0.046}_{-0.044}$ \\
\hspace{1mm}\\                                                                                                                                                                                               
                                                                                                                                                                                                             
$\Omega_{\textrm{m}}$    & $0.369\,^{+0.070}_{-0.065}$ & $0.314\,^{+0.045}_{-0.039}$ & $0.308\,^{+0.016}_{-0.015}$ & $0.304\,^{+0.016}_{-0.014}$ & $0.302\,^{+0.016}_{-0.015}$ & $0.304\,^{+0.016}_{-0.015}$ & $0.304\,^{+0.016}_{-0.016}$ \\
\hspace{1mm}\\                                                                        

$P_{s,1}$                & $<8.13$                     & $<8.17$                     & $<7.91$                     & $<8.06$                     & $<7.85$                     & $<8.09$                     & $<8.11$    \\
\hspace{1mm}\\                                                                        

$P_{s,2}$                & $1.09\,^{+0.42}_{-0.35}$    & $1.01\,^{+0.43}_{-0.35}$    & $1.01\,^{+0.40}_{-0.32}$    & $0.99\,^{+0.42}_{-0.33}$    & $1.02\,^{+0.43}_{-0.34}$    & $1.01\,^{+0.42}_{-0.33}$    & $1.05\,^{+0.43}_{-0.38}$    \\
\hspace{1mm}\\                                                                                                                                                                                               
                                                                                                                                                                                                             
$P_{s,3}$                & $0.68\,^{+0.39}_{-0.36}$    & $0.71\,^{+0.39}_{-0.39}$    & $0.71\,^{+0.39}_{-0.37}$    & $0.72\,^{+0.39}_{-0.38}$    & $0.69\,^{+0.39}_{-0.37}$    & $0.70\,^{+0.40}_{-0.38}$    & $0.69\,^{+0.40}_{-0.39}$    \\
\hspace{1mm}\\                                                                                                                                                                                               
                                                                                                                                                                                                             
$P_{s,4}$                & $1.14\,^{+0.24}_{-0.22}$    & $1.15\,^{+0.24}_{-0.22}$    & $1.15\,^{+0.23}_{-0.21}$    & $1.15\,^{+0.23}_{-0.20}$    & $1.15\,^{+0.23}_{-0.21}$    & $1.15\,^{+0.23}_{-0.21}$    & $1.15\,^{+0.22}_{-0.21}$    \\
\hspace{1mm}\\                                                                                                                                                                                               
                                                                                                                                                                                                             
$P_{s,5}$                & $1.02\,^{+0.11}_{-0.10}$    & $1.01\,^{+0.11}_{-0.11}$    & $1.00\,^{+0.11}_{-0.10}$    & $1.00\,^{+0.11}_{-0.10}$    & $0.99\,^{+0.11}_{-0.10}$    & $0.99\,^{+0.11}_{-0.10}$    & $0.99\,^{+0.11}_{-0.11}$    \\
\hspace{1mm}\\                                                                                                                                                                                               
                                                                                                                                                                                                             
$P_{s,6}$                & $1.03\,^{+0.08}_{-0.07}$    & $1.00\,^{+0.08}_{-0.07}$    & $1.00\,^{+0.08}_{-0.07}$    & $1.00\,^{+0.08}_{-0.07}$    & $0.98\,^{+0.07}_{-0.06}$    & $0.98\,^{+0.07}_{-0.07}$    & $0.98\,^{+0.08}_{-0.07}$    \\
\hspace{1mm}\\                                                                                                                                                                                               
                                                                                                                                                                                                             
$P_{s,7}$                & $0.99\,^{+0.07}_{-0.06}$    & $0.98\,^{+0.08}_{-0.07}$    & $0.98\,^{+0.07}_{-0.07}$    & $0.98\,^{+0.08}_{-0.07}$    & $0.96\,^{+0.07}_{-0.06}$    & $0.95\,^{+0.07}_{-0.06}$    & $0.96\,^{+0.07}_{-0.06}$    \\
\hspace{1mm}\\                                                                                                                                                                                               
                                                                                                                                                                                                             
$P_{s,8}$                & $0.94\,^{+0.06}_{-0.06}$    & $0.95\,^{+0.08}_{-0.07}$    & $0.95\,^{+0.07}_{-0.06}$    & $0.95\,^{+0.08}_{-0.07}$    & $0.94\,^{+0.07}_{-0.06}$    & $0.94\,^{+0.07}_{-0.06}$    & $0.94\,^{+0.07}_{-0.06}$    \\
\hspace{1mm}\\                                                                                                                                                                                               
                                                                                                                                                                                                             
$P_{s,9}$                & $0.92\,^{+0.06}_{-0.05}$    & $0.94\,^{+0.08}_{-0.06}$    & $0.94\,^{+0.07}_{-0.06}$    & $0.94\,^{+0.08}_{-0.06}$    & $0.93\,^{+0.07}_{-0.06}$    & $0.93\,^{+0.07}_{-0.06}$    & $0.94\,^{+0.07}_{-0.06}$    \\
\hspace{1mm}\\                                                                                                                                                                                               
                                                                                                                                                                                                             
$P_{s,10}$               & $0.90\,^{+0.06}_{-0.06}$    & $0.91\,^{+0.08}_{-0.07}$    & $0.91\,^{+0.07}_{-0.06}$    & $0.91\,^{+0.08}_{-0.06}$    & $0.90\,^{+0.07}_{-0.06}$    & $0.90\,^{+0.07}_{-0.06}$    & $0.90\,^{+0.07}_{-0.07}$    \\
\hspace{1mm}\\                                                                                                                                                                                               
                                                                                                                                                                                                             
$P_{s,11}$               & $1.25\,^{+0.30}_{-0.28}$    & $1.24\,^{+0.32}_{-0.31}$    & $1.23\,^{+0.31}_{-0.31}$    & $1.24\,^{+0.31}_{-0.31}$    & $1.22\,^{+0.30}_{-0.31}$    & $1.22\,^{+0.32}_{-0.28}$    & $1.23\,^{+0.31}_{-0.30}$    \\
\hspace{1mm}\\                                                                        

$P_{s,12}$               & {\rm {Unconstrained}}       & {\rm {Unconstrained}}       & {\rm {Unconstrained}}       & {\rm {Unconstrained}}       & {\rm {Unconstrained}}       & {\rm {Unconstrained}}       & {\rm {Unconstrained}}       \\
\hline
\hline
\end{tabular}
\caption{$95\%$~CL constraints on the physical cold dark matter density $\Omega_{\textrm{c}}h^2$, 
the axion mass $m_a$ (in eV), the clustering parameter $\sigma_8$, the relative matter energy density $\Omega_{\textrm{m}}$ and the $P_{s,j}$ parameters for the PPS nodes from the different combinations of data sets explored here in the $\Lambda$CDM+$m_a$ model, considering the \pchip  PPS modeling.}
\label{tab:lcdm+ma+pchip}
\end{center}
\end{table*}

\begin{table*}
\begin{center}\footnotesize
\begin{tabular}{lcccccccc}
\hline \hline
                         & CMB                         & CMB+HST                     & CMB+BAO                     & CMB+BAO                     & CMB+BAO                     & CMB+BAO                     & CMB+BAO\\
                         &                             &                             &                             &    +HST                     & HST+CFHT                    &  +HST+PSZ   (fixed bias)               &  +HST+PSZ\\  
\hline
\hspace{1mm}\\

$\Omega_{\textrm{c}}h^2$ & $0.124\,^{+0.006}_{-0.005}$ & $0.124\,^{+0.005}_{-0.005}$ & $0.122\,^{+0.004}_{-0.004}$ & $0.121\,^{+0.004}_{-0.004}$ & $0.120\,^{+0.003}_{-0.003}$ & $0.119\,^{+0.003}_{-0.003}$ & $0.120\,^{+0.003}_{-0.003}$  \\
\hspace{1mm}\\           
                         
$m_a$ [eV]               & $<1.83$                     & $<1.56$                     & $<0.84$                     & $<0.83$                     & $<1.16$                     & $0.80\,^{+0.53}_{-0.50}$    & $<1.26$                      \\
\hspace{1mm}\\           
                         
$\sigma_8$               & $0.785\,^{+0.064}_{-0.083}$ & $0.791\,^{+0.057}_{-0.076}$ & $0.803\,^{+0.041}_{-0.048}$ & $0.803\,^{+0.041}_{-0.048}$ & $0.783\,^{+0.047}_{-0.054}$ & $0.758\,^{+0.028}_{-0.029}$ & $0.767\,^{+0.045}_{-0.045}$  \\
\hspace{1mm}\\           
                         
$\Omega_{\textrm{m}}$    & $0.337\,^{+0.048}_{-0.044}$ & $0.328\,^{+0.041}_{-0.039}$ & $0.310\,^{+0.025}_{-0.023}$ & $0.308\,^{+0.024}_{-0.023}$ & $0.305\,^{+0.025}_{-0.024}$ & $0.307\,^{+0.027}_{-0.026}$ & $0.306\,^{+0.027}_{-0.025}$  \\
\hspace{1mm}\\           
                         
$\log[10^{10} A_s]$ & $3.10\,^{+0.05}_{-0.05}$    & $3.10\,^{+0.05}_{-0.05}$    & $3.10\,^{+0.05}_{-0.05}$    & $3.10\,^{+0.05}_{-0.05}$    & $3.10\,^{+0.05}_{-0.05}$    & $3.09\,^{+0.05}_{-0.05}$    & $3.09\,^{+0.05}_{-0.05}$     \\
\hspace{1mm}\\           
                         
$n_s$                    & $0.961\,^{+0.014}_{-0.015}$ & $0.963\,^{+0.013}_{-0.014}$ & $0.968\,^{+0.011}_{-0.011}$ & $0.969\,^{+0.011}_{-0.011}$ & $0.971\,^{+0.011}_{-0.011}$ & $0.973\,^{+0.011}_{-0.011}$ & $0.972\,^{+0.011}_{-0.011}$  \\
\hline
\hline
\end{tabular}
\caption{$95\%$~CL constraints on 
$\Omega_{\textrm{c}}h^2$, 
the axion mass $m_a$ (in eV), 
$\sigma_8$, $\Omega_{\textrm{m}}$, ${\rm{log}}(10^{10} A_s)$ and $n_s$
from the different combinations of data sets explored here in the $\Lambda$CDM+$m_a$ model, assuming the standard power-law PPS.}
\label{tab:lcdm+ma+pl}
\end{center}
\end{table*}

\begin{figure*}[!t]
\includegraphics[width=11.1cm]{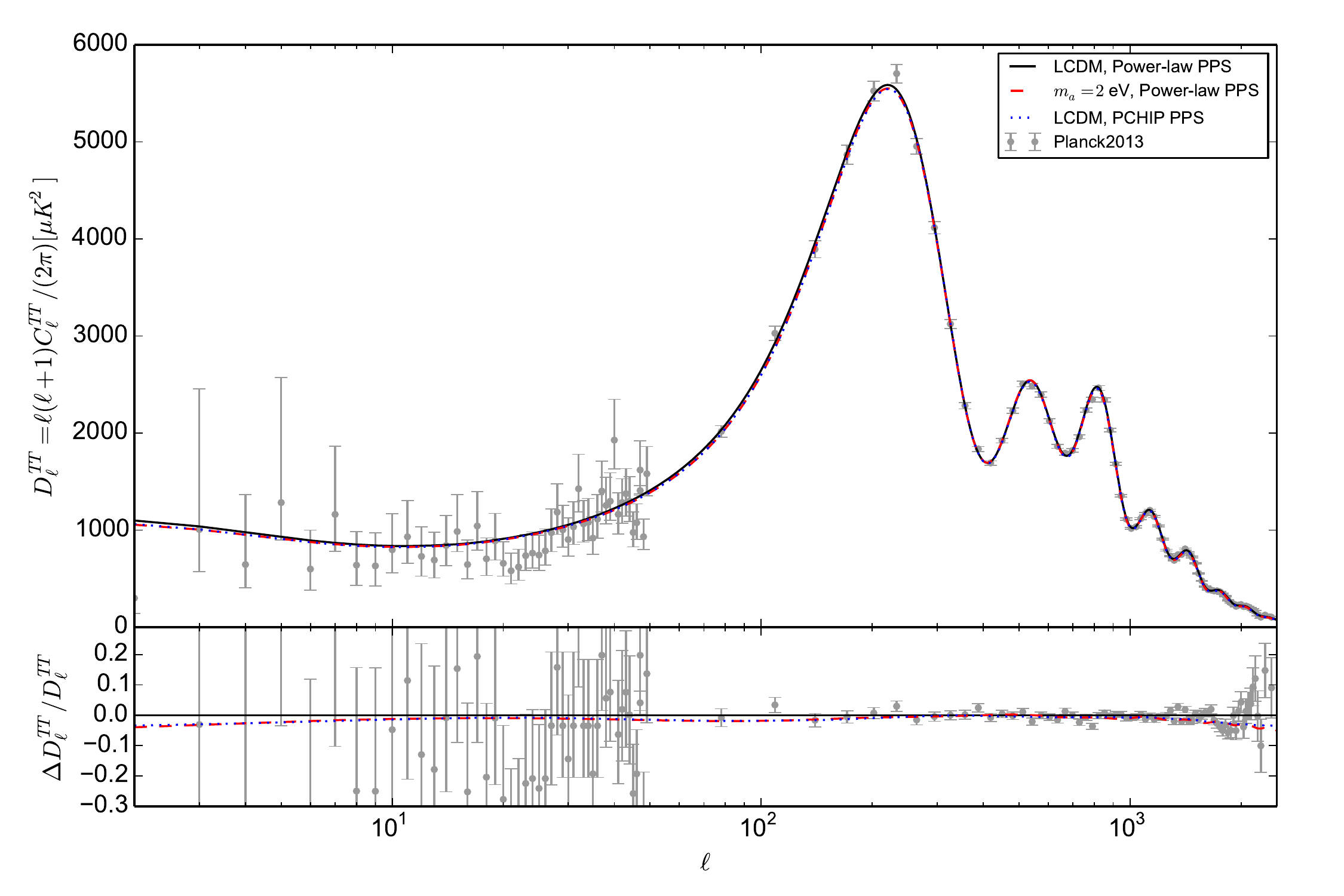}
 \caption{Temperature anisotropies for the pure $\Lambda$CDM model and a power-law PPS (solid red line), for a $2$~eV thermal axion
mass and a power-law PPS (dashed blue line) and for the standard $\Lambda$CDM model but the PPS described by the \pchip model (dotted black line). The data points and the error bars in the left panel show the measurements of the photon temperature anisotropies arising from the Planck 2013 data release~\cite{Ade:2013zuv}.}
\label{fig:plottt}
\end{figure*}

Table \ref{tab:lcdm+ma+pchip} depicts our results in the first scenario explored
here, in which the axion mass is a free parameter and the PPS is
described by the approach specified in Sec.~\ref{sub:pps}. Concerning CMB measurements only, the
bounds on the thermal axion masses are largely relaxed in the case in which the PPS
is not described by a simple power-law, as can be noticed after comparing the results depicted in Tab.~\ref{tab:lcdm+ma+pchip} with those shown in Tab.~\ref{tab:lcdm+ma+pl}.  This can be understood in terms of Fig.~\ref{fig:plottt}, which
illustrates the degeneracies in the temperature anisotropies between the
thermal axion mass and the \pchip PPS.
Figure~\ref{fig:plottt} shows the temperature anisotropies for a $\Lambda$CDM model and a power-law PPS (solid red line), for a $2$~eV thermal axion
mass and a power-law PPS (dashed blue line) and  for a $\Lambda$CDM model
but the PPS described by the \pchip model explored here (dotted black line), with values
for the $P_{s,j}$ chosen to match the non-zero thermal axion mass curve, accordingly to their current allowed regions 
(see Tab.~\ref{tab:lcdm+ma+pchip}). More concretely, we have used the following values for the PPS parameters: 
$P_{s,1}=1.15$, $P_{s,2}=1.073$, $P_{s,3}=1.058$, $P_{s,4}=1.03$, $P_{s,5}=0.99$, $P_{s,6}=0.97$,
$P_{s,7}=0.966$, $P_{s,8}=0.932$, $P_{s,9}=0.91$, $P_{s,10}=0.86$, $P_{s,11}=0.84$ and $P_{s,12}=0.77$.
Notice that the case of a $2$~eV thermal axion can be easily mimicked
by a simple $\Lambda$CDM model if the assumptions concerning the PPS 
shape are relaxed. We also add in this figure the measurements of the photon temperature anisotropies from the Planck 2013 data release~\cite{Ade:2013zuv}.

The addition to the CMB data of the HST prior on the Hubble constant
provides a $95\%$~CL upper limit on the thermal axion mass of
$1.31$~eV~\footnote{There exists a very large degeneracy between $H_0$
  and the neutrino masses when restricting the numerical analyses to
  CMB measurements. The addition of the HST prior on the Hubble
  constant helps enormously in breaking this degeneracy,
  see~\cite{Giusarma:2012ph}.}, while the further addition of the BAO
measurements brings this constraint down to $0.91$~eV, as these last
data sets are directly sensitive to the free-streaming nature of the
thermal axion. Notice that these two $95\%$~CL upper bounds are very
similar to the ones obtained when considering the standard power-law
power spectrum, which are $1.56$~eV and $0.83$~eV for the CMB+HST and
CMB+HST+BAO data combinations, respectively.

Interestingly, when adding the CFHT bounds on the $\sigma_8$-$\Omega_m$ relationship, the bounds on the thermal axion mass become weaker. 
The reason for that is due to the lower $\sigma_8$ values preferred by weak lensing measurements, 
values that can be achieved by allowing for higher axion masses. 
The larger the axion mass, the larger is the reduction of the matter power spectrum at small (i.e. cluster) scales, 
leading consequently to a smaller value of the clustering parameter $\sigma_8$.  

If we instead consider now the PSZ data set with fixed cluster mass
bias, together with the CMB, BAO and HST measurements, a non-zero
value of the thermal axion mass of $\sim 1$~eV ($\sim 0.80$~eV) is
favoured at $\sim4\sigma$ ($\sim3\sigma$) level, when considering the
\pchip (standard power-law) PPS approach~\footnote{A similar effect when considering PSZ data for constraining either thermal axion or neutrino masses has also been found in Refs.~\cite{Hamann:2013iba,Wyman:2013lza,Giusarma:2014zza,Dvorkin:2014lea,Archidiacono:2014apa}.}. However, these results must be regarded as an illustration of what could be achieved with future cluster mass calibrations, as the Planck collaboration has recently shown in their analyses of the 2015 Planck cluster catalogue~\cite{Ade:2015fva}. When more realistic approaches for the cluster mass bias are used, the errors on the so-called cluster normalization condition are larger, and, consequently, the preference for a non-zero axion mass  of $1$~eV is only mild in the \pchip PPS case, while in the case of a standard power-law PPS such an evidence completely disappears.

\begin{figure*}[!t]
\begin{tabular}{c c}
\includegraphics[width=8.3cm]{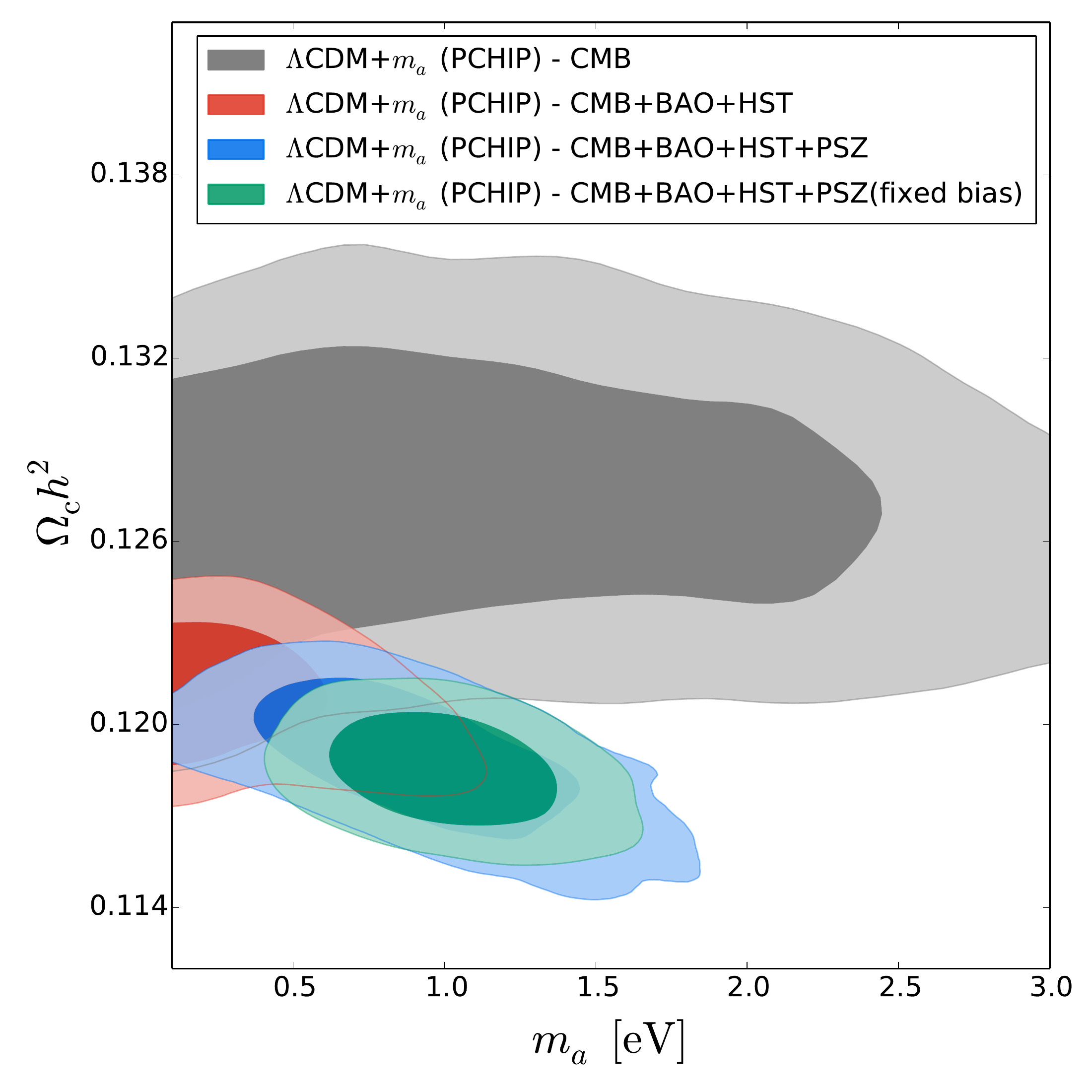}&\includegraphics[width=8.3cm]{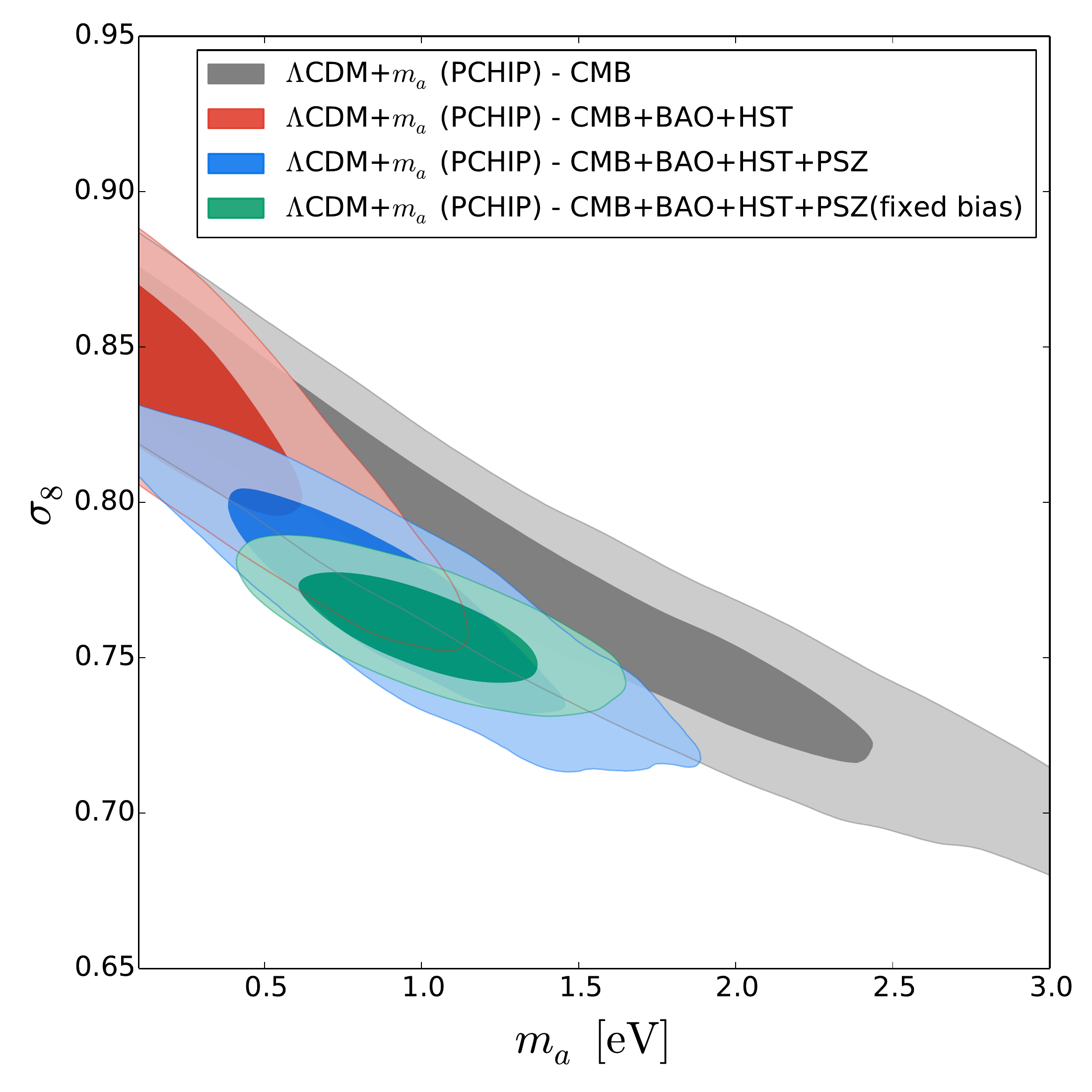}\\
\end{tabular}
 \caption{The left panel depicts the $68\%$ and $95\%$~CL allowed
   regions in the ($m_a$, $\Omega_c h^2$) plane for different possible
   data combinations, when a \pchip{} PPS is assumed. The right panel shows the equivalent but in the
   ($m_a$, $\sigma_8$) plane.}
\label{fig:ma_pchip}
\end{figure*}

Figure~\ref{fig:ma_pchip} (left panel) shows the $68\%$ and $95\%$~CL
allowed regions in the ($m_a$, $\Omega_c h^2$) plane for some of the
possible data combinations explored in this study, and assuming the \pchip PPS modeling. Notice that, when
adding BAO measurements,  lower values of the physical cold dark matter density are
preferred. This is due to the fact that large scale structure allows
for lower axion masses than CMB data alone. 
The lower is the thermal axion mass,
the lower is the amount of hot dark matter and consequently the lower should be
the cold dark matter component. 
This effect is clear 
from the results shown in Tab.~\ref{tab:lcdm+ma+pchip} and Tab.~\ref{tab:lcdm+ma+pl}, where the values of the
physical cold dark matter  density $\Omega_c h^2$ and of the relative
current matter density $\Omega_m$ arising from our numerical fits
are shown, for the different data combinations considered here. 

The right panel of Fig.~\ref{fig:ma_pchip} shows the $68\%$ and $95\%$~CL
allowed regions in the ($m_a$, $\sigma_8$) plane in the \pchip PPS scenario. The lower values of
the $\sigma_8$ clustering parameter preferred by PSZ data (see the results shown in Tab.~\ref{tab:lcdm+ma+pchip} and Tab.~\ref{tab:lcdm+ma+pl}) are translated into a preference for non-zero thermal axion masses. Larger values of $m_a$ will enhance the matter power spectrum suppression at scales below the axion free-streaming scale, leading to smaller values of the $\sigma_8$ clustering parameter, as preferred by PSZ measurements.  The evidence for non-zero axion masses is more significant when fixing the cluster mass bias in the PSZ data analyses. 

\begin{figure*}[!t]
\begin{tabular}{c c}
\includegraphics[width=8.3cm]{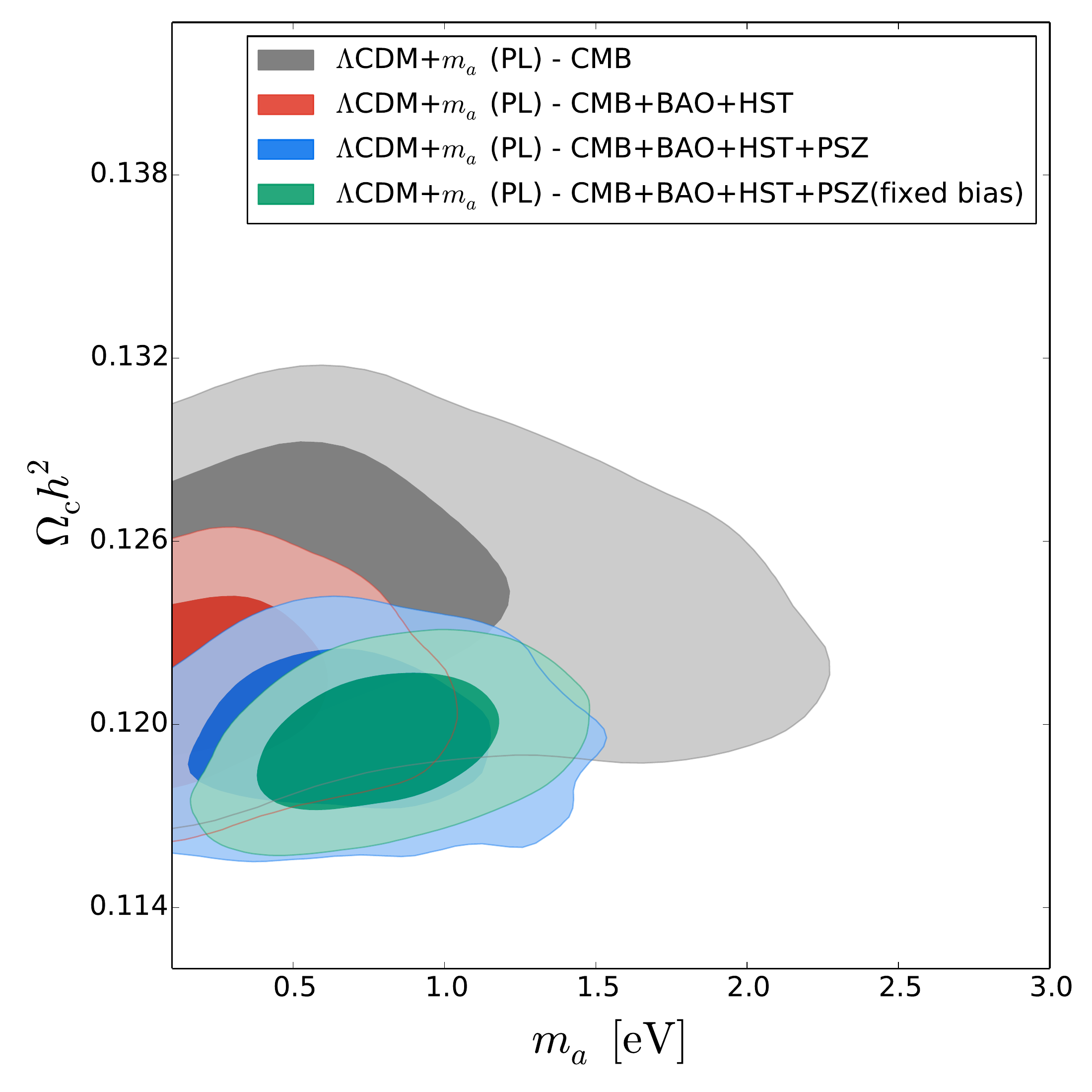}&\includegraphics[width=8.3cm]{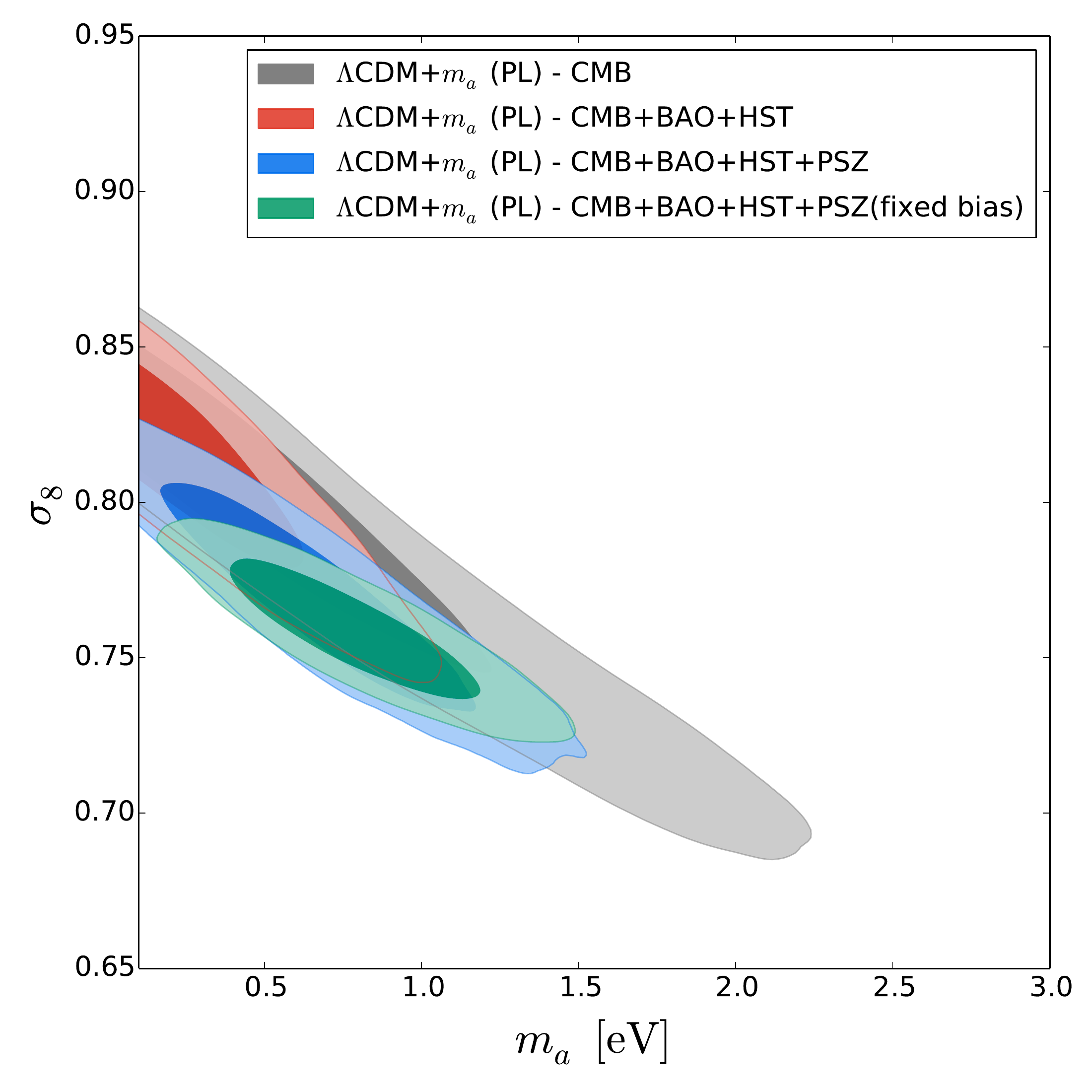}\\
\end{tabular}
 \caption{The left panel depicts the $68\%$ and $95\%$~CL allowed
   regions in the ($m_a$, $\Omega_c h^2$) plane for different possible
   data combinations, when a power-law PPS is assumed. The right panel shows the equivalent but in the
   ($m_a$, $\sigma_8$) plane.}
\label{fig:ma_pl}
\end{figure*}

Figure \ref{fig:ma_pl} shows the equivalent to Fig.~\ref{fig:ma_pchip} but for a standard power-law PPS. Notice that, except for the case in which CMB measurements are considered alone, the thermal axion mass constraints do not change significantly, if they are compared to the \pchip PPS modeling. 
This fact clearly states the robustness of the cosmological bounds on thermal axion masses and it is applicable to the remaining cosmological parameters, see Tabs.~\ref{tab:lcdm+ma+pchip} and \ref{tab:lcdm+ma+pl}. Note that, for the standard case of a power-law PPS, the preference for non-zero axion masses appears only when considering the (unrealistic) PSZ analysis with a fixed cluster mass bias. When more realistic PSZ measurements of the cluster normalization condition  are exploited, there is no preference for a non-zero thermal axion mass. 


\begin{figure}[!t]
\includegraphics[width=9cm]{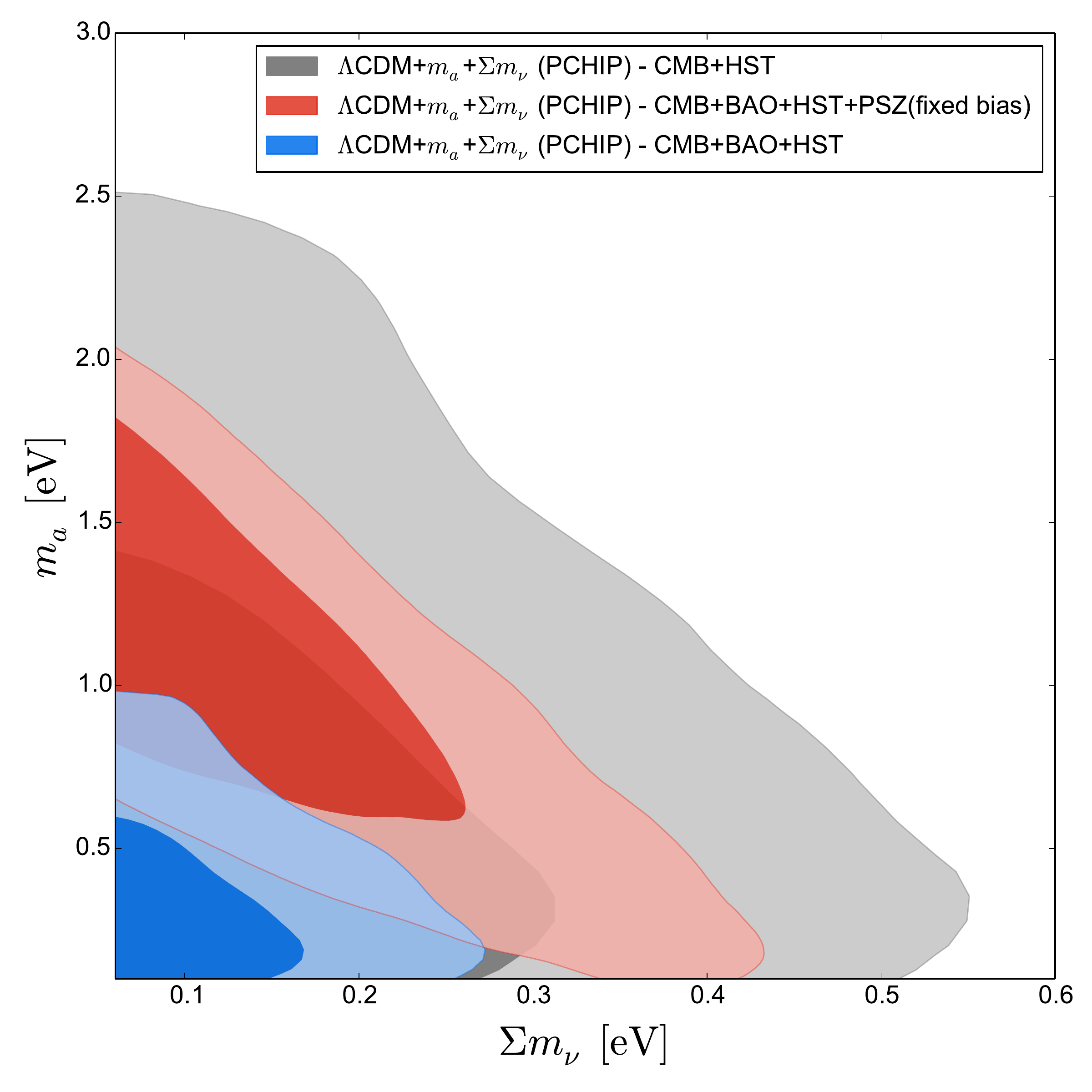}
\caption{$68\%$ and $95\%$~CL allowed
   regions in the ($\sum m_\nu$, $m_a$) plane, both in eV, for three different possible
   data combinations, when a \pchip{} PPS is assumed.}
\label{fig:mamnu}
\end{figure}
The last scenario we explore here is a $\Lambda$CDM+$m_a$+$\sum m_\nu$ universe, in which we consider two coexisting hot dark matter species: thermal axions and three active (massive) neutrinos.
Table~\ref{tab:lcdm+ma+mnu+pchip} illustrates the equivalent of Tab.~\ref{tab:lcdm+ma+pchip} but including the active neutrino masses in the MCMC parameters. We do not perform here the analysis for the hot mixed dark matter model with the standard power-law matter power spectrum, as it was already presented previously in Ref.~\cite{Giusarma:2014zza}.  If we compare to the standard power-law case, we find that the bounds on the axion and neutrino masses presented here are very similar. Furthermore, no evidence for neutrino masses nor for a non-zero axion mass appears in this mixed hot dark matter scenario (except for the axion case and only if considering PSZ clusters with the bias fixed). The reason for that is due to the strong degeneracy between $m_a$ and $\sum m_\nu$, see Fig.~\ref{fig:mamnu}, where one can notice that that these two parameters are negatively correlated: an increase in the axion mass will increase the amount of the hot dark matter component. In order to compensate the changes in both the CMB temperature anisotropies (via the early ISW effect) and in the power spectrum (via the suppression at small scales of galaxy clustering),  the contribution to the hot dark matter from the neutrinos should be reduced. We have shown in Fig.~\ref{fig:mamnu} three possible data combinations. Notice that for the case in which PSZ cluster measurements (with the bias fixed) are included the strong degeneracy between $m_a$ and $\sum m_\nu$ is partially broken, due to the smaller value of $\sigma_8$ preferred by the former data set. However,  these results strongly rely on the numerical results concerning the cluster mass bias and therefore the evidence for $m_a\neq 0$ should be regarded as what could be obtained in if these measurements are further supported by independent data from future cluster surveys.

\begin{table*}
\begin{center}\footnotesize
\begin{tabular}{lcccccccc}
\hline \hline
                         & CMB                         & CMB+HST                     & CMB+BAO                     & CMB+BAO                     & CMB+BAO                     & CMB+BAO                      & CMB+BAO                       \\
                         &                             &                             &                             &    +HST                     & +HST+CFHT                   &  +HST+PSZ(fixed bias)        &  +HST+PSZ                     \\  
\hline
\hspace{1mm}\\

$\Omega_{\textrm{c}}h^2$ & $0.130\,^{+0.008}_{-0.007}$ & $0.125\,^{+0.006}_{-0.007}$ & $0.121\,^{+0.003}_{-0.003}$ & $0.121\,^{+0.003}_{-0.003}$ & $0.119\,^{+0.003}_{-0.003}$ & $0.118\,^{+0.003}_{-0.003}$  & $0.118\,^{+0.003}_{-0.003}$  \\
\hspace{1mm}\\                                                                                                                                                                  
                                                                                                                                                                                
$m_a$ [eV]               & $<2.48$                     & $<1.64$                     & $<0.81$                     & $<0.86$                     & $<1.23$                     & $0.81\,^{+0.59}_{-0.69}$     & $<1.46$                      \\
\hspace{1mm}\\                                                                                                                                                                  
                                                                                                                                                                                
$\sum m_\nu$ [eV]        & $<2.11$                     & $<0.43$                     & $<0.22$                     & $<0.21$                     & $<0.27$                     & $<0.32$                      & $<0.35$                      \\
\hspace{1mm}\\                                                                                                                                                                  
                                                                                                                                                                                
$\sigma_8$               & $0.700\,^{+0.172}_{-0.202}$ & $0.803\,^{+0.082}_{-0.091}$ & $0.833\,^{+0.055}_{-0.058}$ & $0.834\,^{+0.058}_{-0.064}$ & $0.787\,^{+0.052}_{-0.055}$ & $0.766\,^{+0.043}_{-0.044}$  & $0.757\,^{+0.023}_{-0.022}$  \\
\hspace{1mm}\\                                                                                                                                                                  
                                                                                                                                                                                
$\Omega_{\textrm{m}}$    & $0.486\,^{+0.277}_{-0.193}$ & $0.356\,^{+0.064}_{-0.062}$ & $0.309\,^{+0.016}_{-0.015}$ & $0.308\,^{+0.016}_{-0.015}$ & $0.306\,^{+0.015}_{-0.015}$ & $0.308\,^{+0.016}_{-0.016}$  & $0.308\,^{+0.017}_{-0.016}$  \\
\hspace{1mm}\\                                                                                                                                                                  
                                                                                                                                                                                
$P_{s,1}$                & $<8.01$                     & $<8.13$                     & $<7.00$                     & $<8.17$                     & $<7.59$                     & $<8.29$                      & $<8.18$                      \\
\hspace{1mm}\\                                                                                                                                                                  
                                                                                                                                                                                
$P_{s,2}$                & $1.17\,^{+0.42}_{-0.38}$    & $1.09\,^{+0.41}_{-0.37}$    & $1.03\,^{+0.40}_{-0.35}$    & $1.02\,^{+0.39}_{-0.34}$    & $1.02\,^{+0.40}_{-0.32}$    & $1.03\,^{+0.36}_{-0.34}$     & $1.05\,^{+0.40}_{-0.36}$     \\
\hspace{1mm}\\                                                                                                                                                                  
                                                                                                                                                                                
$P_{s,3}$                & $0.66\,^{+0.37}_{-0.35}$    & $0.69\,^{+0.38}_{-0.37}$    & $0.70\,^{+0.38}_{-0.38}$    & $0.72\,^{+0.38}_{-0.37}$    & $0.68\,^{+0.37}_{-0.33}$    & $0.71\,^{+0.40}_{-0.39}$     & $0.69\,^{+0.39}_{-0.37}$     \\
\hspace{1mm}\\                                                                                                                                                                  
                                                                                                                                                                                
$P_{s,4}$                & $1.17\,^{+0.23}_{-0.23}$    & $1.15\,^{+0.23}_{-0.22}$    & $1.15\,^{+0.22}_{-0.21}$    & $1.15\,^{+0.21}_{-0.21}$    & $1.15\,^{+0.20}_{-0.19}$    & $1.14\,^{+0.21}_{-0.20}$     & $1.16\,^{+0.22}_{-0.21}$     \\
\hspace{1mm}\\                                                                                                                                                                  
                                                                                                                                                                                
$P_{s,5}$                & $1.05\,^{+0.15}_{-0.14}$    & $1.01\,^{+0.11}_{-0.10}$    & $1.00\,^{+0.11}_{-0.10}$    & $1.00\,^{+0.11}_{-0.10}$    & $0.98\,^{+0.11}_{-0.10}$    & $0.99\,^{+0.11}_{-0.10}$     & $0.98\,^{+0.11}_{-0.10}$     \\
\hspace{1mm}\\                                                                                                                                                                  
                                                                                                                                                                                
$P_{s,6}$                & $1.04\,^{+0.09}_{-0.08}$    & $1.01\,^{+0.08}_{-0.07}$    & $1.00\,^{+0.07}_{-0.07}$    & $1.00\,^{+0.07}_{-0.07}$    & $0.98\,^{+0.07}_{-0.06}$    & $0.98\,^{+0.07}_{-0.07}$     & $0.98\,^{+0.07}_{-0.07}$     \\
\hspace{1mm}\\                                                                                                                                                                  
                                                                                                                                                                                
$P_{s,7}$                & $0.99\,^{+0.06}_{-0.06}$    & $0.98\,^{+0.07}_{-0.06}$    & $0.98\,^{+0.07}_{-0.07}$    & $0.98\,^{+0.07}_{-0.07}$    & $0.95\,^{+0.07}_{-0.06}$    & $0.95\,^{+0.06}_{-0.06}$     & $0.95\,^{+0.07}_{-0.06}$     \\
\hspace{1mm}\\                                                                                                                                                                  
                                                                                                                                                                                
$P_{s,8}$                & $0.93\,^{+0.06}_{-0.05}$    & $0.94\,^{+0.06}_{-0.06}$    & $0.95\,^{+0.07}_{-0.07}$    & $0.95\,^{+0.07}_{-0.07}$    & $0.93\,^{+0.07}_{-0.05}$    & $0.94\,^{+0.07}_{-0.06}$     & $0.93\,^{+0.07}_{-0.06}$     \\
\hspace{1mm}\\                                                                                                                                                                  
                                                                                                                                                                                
$P_{s,9}$                & $0.91\,^{+0.06}_{-0.05}$    & $0.93\,^{+0.06}_{-0.06}$    & $0.94\,^{+0.07}_{-0.06}$    & $0.94\,^{+0.07}_{-0.06}$    & $0.93\,^{+0.07}_{-0.06}$    & $0.93\,^{+0.06}_{-0.06}$     & $0.93\,^{+0.07}_{-0.06}$     \\
\hspace{1mm}\\                                                                                                                                                                  
                                                                                                                                                                                
$P_{s,10}$               & $0.90\,^{+0.06}_{-0.06}$    & $0.90\,^{+0.07}_{-0.06}$    & $0.91\,^{+0.07}_{-0.07}$    & $0.91\,^{+0.08}_{-0.07}$    & $0.88\,^{+0.07}_{-0.06}$    & $0.89\,^{+0.07}_{-0.07}$     & $0.90\,^{+0.07}_{-0.07}$     \\
\hspace{1mm}\\                                                                                                                                                                  
                                                                                                                                                                                
$P_{s,11}$               & $2.18\,^{+0.85}_{-0.77}$    & $2.07\,^{+0.81}_{-0.80}$    & $2.12\,^{+0.90}_{-0.86}$    & $2.15\,^{+0.95}_{-0.94}$    & $1.64\,^{+0.79}_{-0.75}$    & $1.83\,^{+0.87}_{-0.86}$     & $1.84\,^{+0.86}_{-0.87}$     \\
\hspace{1mm}\\                                          

$P_{s,12}$               & {\rm {Unconstrained}}       & {\rm {Unconstrained}}       & {\rm {Unconstrained}}       & {\rm {Unconstrained}}       & {\rm {Unconstrained}}  &      {\rm {Unconstrained}}         & {\rm {Unconstrained}}       \\
\hline
\hline
\end{tabular}

\caption{$95\%$~CL constraints on the physical cold dark matter density $\Omega_{\textrm{c}}h^2$, 
the axion mass $m_a$, the sum of the active neutrino masses $\sum m_\nu$ (both in eV), the clustering parameter $\sigma_8$, the relative matter energy density $\Omega_{\textrm{m}}$ and the $P_{s,j}$ parameters for the PPS nodes from the different combinations of data sets explored here in the $\Lambda$CDM+$m_a$+$\sum m_\nu$ model, considering the \pchip  PPS modeling.}
\label{tab:lcdm+ma+mnu+pchip}
\end{center}
\end{table*}

\begin{figure*}[!t]
\includegraphics[width=12cm]{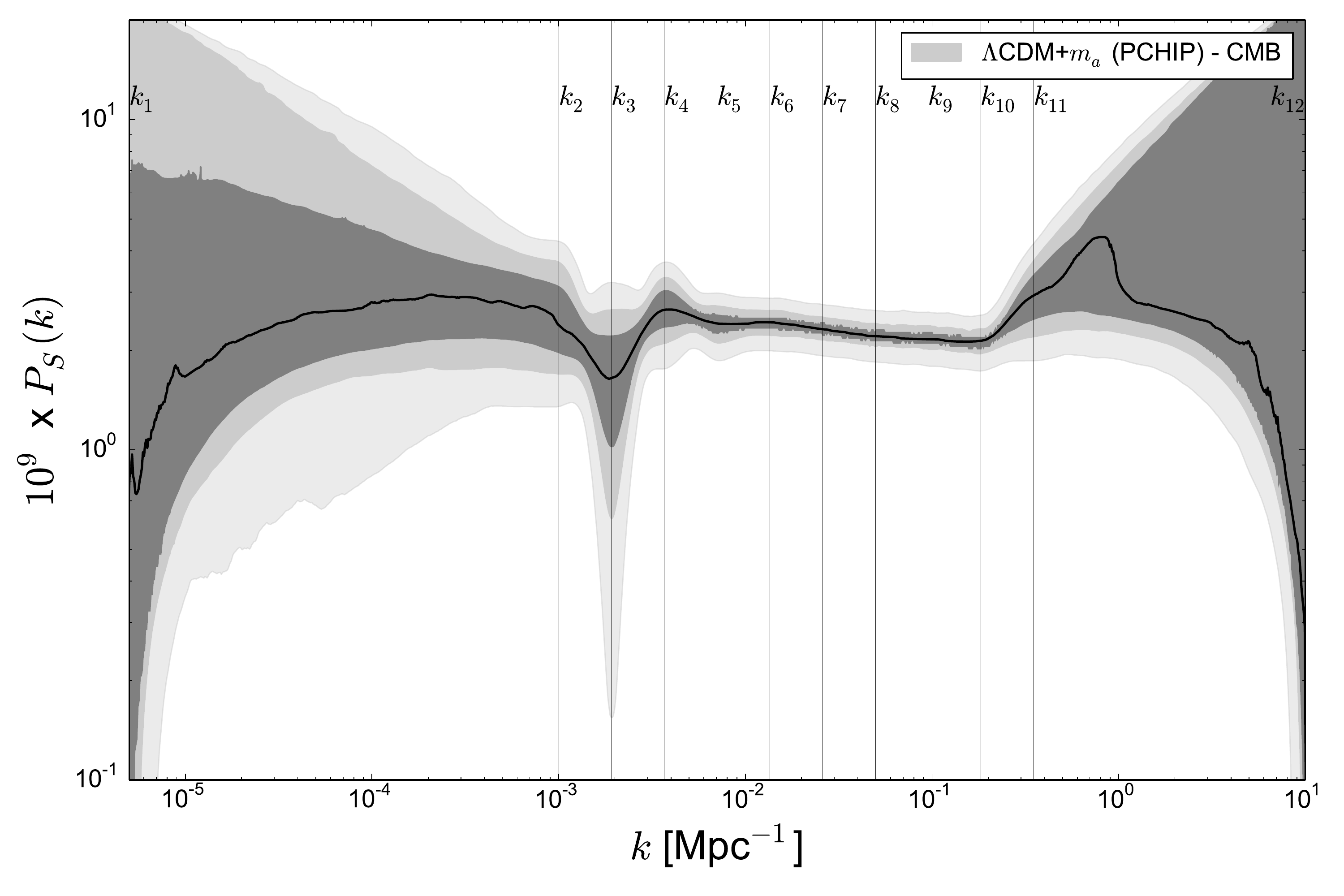}
\caption{$68\%$, $95\%$ and $99\%$~CL allowed regions for the \pchip PPS scale dependence in the $\Lambda$CDM+$m_a$ model, using CMB data only.
The bands are obtained with a marginalization of the posterior distribution for each different value of the wavenumber $k$ in a fine grid.
The black line represents the peak of the posterior distribution at each value of $k$.}
\label{fig:outputPPS}
\end{figure*}

Besides the results concerning the thermal axion mass and the standard $\Lambda$CDM parameters,
we also obtain constraints on the form of the PPS when modeled accordingly to the \pchip scenario.
The 95\% CL limits for the $P_{s,j}$ parameters are shown in Tab.~\ref{tab:lcdm+ma+pchip}, 
while an example of the reconstructed PPS is given in Fig.~\ref{fig:outputPPS}, 
where we show the 68\%, 95\% and 99\% CL allowed regions arising from a fit to CMB data of the \pchip PPS scale dependence, 
 in the context of a $\Lambda$CDM+$m_a$ model.
We do not show the corresponding figures obtained from all the other data combinations 
since they are equivalent to Fig.~\ref{fig:outputPPS}, as one can infer from the very small 
differences in the $95\%$~CL allowed ranges for the $P_{s,j}$ parameters arising from different data sets, see Tab.~\ref{tab:lcdm+ma+pchip}.
Note that both $P_{s,1}$ and $P_{s,12}$ are poorly constrained at this confidence level:
the reason for that is the absence of measurements at their corresponding wavenumbers.
All the remaining $P_{s,j}$, with $j=2,\ldots,11$ are
well-constrained. In particular, in the range between $k_5$
and $k_{10}$ (see Eq.~(\ref{eq:nodesspacing})), the $P_{s,j}$ are determined with few percent accuracy.  Indeed, in the range covered between these nodes, the PPS does not present features and can be perfectly described by a power-law parametrization.
Among the interesting features outside the former range, we can notice  in Fig.~\ref{fig:outputPPS}
a significant dip at wavenumbers around $k=0.002\mpcinv$, that comes from the dip at $\ell=20-30$ in the CMB temperature power spectrum and a small bump around $k=0.0035\mpcinv$, corresponding to the increase at $\ell\simeq40$. These features have been obtained in previous works \cite{Hunt:2013bha,Hazra:2014jwa,Gariazzo:2014dla} using different methods and data sets. In addition, we obtain an increase of power at $k\simeq0.2\mpcinv$, necessary to compensate the effects of the thermal axion mass in both the temperature anisotropies and the large scale structure of the universe.

\section{Conclusions}
\label{sec:concl}
Axions provide the most elegant scenario to solve the strong CP problem, and may be produced in the early universe via both thermal and non-thermal processes. While non thermal axions are highly promising cold dark matter candidates,  their thermal companions will contribute to the hot dark matter component of the universe, together with the (light) three active neutrinos of the standard model of elementary particles. Therefore, the cosmological consequences of light massive thermal axions are very much alike those associated with neutrinos, as axions also have a free-streaming nature, suppressing structure formation at small scales. Furthermore, these light thermal axions will also contribute to the dark radiation background, leading to deviations of the  relativistic degrees of freedom $\neff$ from its canonically expected value of $\neff=3.046$. Based on these signatures, several studies have been carried out in the literature deriving bounds on the thermal axion mass~\cite{Melchiorri:2007cd,Hannestad:2007dd,Hannestad:2008js,Hannestad:2010yi,Archidiacono:2013cha,Giusarma:2014zza}. 

Nevertheless, these previous constraints assumed that the underlying
primordial perturbation power spectrum follows the usual power-law
description governed, in its most economical form, by an amplitude
and a scalar spectral index. Here we have relaxed such an assumption, 
in order to test the robustness of the cosmological axion mass
bounds. Using an alternative, non-parametric description of the
primordial power spectrum of the scalar perturbations,
named \pchip and introduced in Ref.~\cite{Gariazzo:2014dla}, we have shown that, in practice, when combining CMB measurements with low redshift cosmological probes, the axion mass constraints are only mildly sensitive to the primordial
power spectrum choice and therefore are not strongly dependent on the
particular details of the underlying inflationary model. 
These results agree with the findings of Ref.~\cite{dePutter:2014hza} for the neutrino mass case. The tightest
 bound we find in the \pchip primordial power spectrum approach is obtained when considering BAO measurements together with CMB data, with $m_a<
 0.89$~eV at $95\%$~CL. In the standard power-law primordial power
 spectrum modeling, the tightest bound is $m_a<
 0.83$~eV at $95\%$~CL, obtained when combining BAO, CMB and HST measurements. Notice that these
 bounds are very similar, confirming the robustness of the cosmological
 axion mass measurements versus the primordial power spectrum
 modeling. 

Interestingly, both weak lensing measurements and cluster number
counts weaken the thermal axion mass bounds. The reason for that is
due to the lower $\sigma_8$ values preferred by 
these measurements, which could be generated by a larger axion
mass. More concretely, Planck cluster measurements provide a
measurement of the so-called cluster normalization condition,
which establishes a relationship between the clustering parameter
$\sigma_8$ and the current matter mass-energy density $\Omega_m$. 
However, the errors  on this relationship depend crucially on the
knowledge of the cluster mass bias. A conservative approach for the
cluster mass calibration results in a mild (zero) evidence for a
non-zero axion mass  of $1$~eV in the \pchip (power-law) PPS case.
We also illustrate a case in which the cluster mass bias is fixed, to forecast the expected results from future cosmological
measurements. In this case, a non-zero
value of the thermal axion mass of $\sim 1$~eV ($\sim 0.80$~eV) is
favoured at $\sim4\sigma$ ($\sim3\sigma$) level, when considering the
\pchip (power-law) PPS approach. 
When considering additional hot relics in our analyses, as the sum of the three active neutrino masses, the evidence for 
a $\sim 1$~eV thermal axion mass disappears almost completely. Furthermore, these values of axion masses correspond to an axion coupling constant 
$f_a= 6\times 10^6$~GeV, which seems to be in tension with the limits extracted from the neutrino signal duration from SN 1987A~\cite{Raffelt:2006cw,Raffelt:1990yz} (albeit these limits depend strongly on the precise axion emission rate and still remain rough estimates). Precise cluster mass calibration measurements are therefore mandatory to assess whether there exists a cosmological indication for non-zero axion masses, as the cluster mass bias is highly correlated with the clustering parameter $\sigma_8$, which, in turn, is highly affected by the free-streaming nature of a hot dark matter component, as thermal axions.

\section{Acknowledgments}
OM is supported by PROMETEO II/2014/050, by the Spanish Grant FPA2011--29678 of the MINECO and by PITN-GA-2011-289442-INVISIBLES. This work has been done within the Labex ILP (reference ANR-10-LABX-63) part of the Idex SUPER, and received financial state aid managed by the Agence Nationale de la Recherche, as part of the programme Investissements d'avenir under the reference ANR-11-IDEX-0004-02. EDV acknowledges the support of the European Research Council via the Grant  number 267117 (DARK, P.I. Joseph Silk).



\end{document}